\documentclass[a4paper,11pt]{article}
\usepackage{jcappub}
\usepackage{amssymb}
\usepackage{graphicx}
\usepackage{amsmath}
\usepackage{hyperref}
\usepackage{subcaption}
\usepackage{slashed}
\usepackage{xcolor}
\usepackage{orcidlink}

\newcommand{\md}{\mathrm{d}}
\newcommand{\Tr}{\operatorname{Tr}}

\newcommand{\Det}{\operatorname{Det}}
\newcommand{\PBH}{\mathrm{PBH}}

\usepackage{tikz-feynman}
\tikzfeynmanset{compat=1.1.0}

\title{
Chiral phase transition with primordial black holes: 
Distinct phase structure and catalysis
}

\author{Masanori Tanaka\,\orcidlink{https://orcid.org/0000-0002-1303-7043}\,$^{a,1}$, Jun-Chen Wang\,\orcidlink{https://orcid.org/0009-0004-2226-3183}\,$^{b,1}$, Jing-Jun Zhang\,\orcidlink{https://orcid.org/0009-0007-5228-5959}\,$^{b,1}$}

\affiliation{$^{a}$ Center for High Energy Physics, Peking University, Beijing 100871, China}

\affiliation{$^{b}$ School of Physics, Peking University, Beijing 100871, China}

\abstract{
We study the impact of primordial black holes (PBHs) on the chiral phase transition and its associated stochastic gravitational-wave (GW) signals. Using the three-flavor Nambu-Jona-Lasinio model, we construct the chiral effective potential in a Schwarzschild spacetime background. We find that PBHs promote chiral symmetry restoration and induce a nontrivial local phase structure in the vicinity of the event horizon simultaneously. In particular, this structure exhibits a novel chiral symmetry breaking pattern involving both second- and first-order phase transitions, a feature absent in flat spacetime.
We further demonstrate that PBHs act as genuine catalysts for the chiral phase transition by analyzing the bounce solution in curved spacetime. The presence of PBHs substantially enhances the inverse duration parameter $\beta/H$ while leaving the overall transition strength comparable to that in flat spacetime. As a consequence, even a small population of PBHs can induce $\mathcal{O}(1)$ shifts in both the peak frequency and the peak amplitude of the GW spectrum generated by the first-order dark chiral phase transition. 
\note{All authors contributed equally to this work}
}

\emailAdd{tanaka@pku.edu.cn}

\emailAdd{junchenwang@stu.pku.edu.cn}

\emailAdd{zhang\_jingjun@stu.pku.edu.cn}

\begin{document}

\maketitle

\section{Introduction}

Cosmological first-order phase transitions (FOPTs) are intriguing phenomena that may have occurred in the early universe.
During FOPTs, vacuum bubbles separating false and true vacua can nucleate stochastically via quantum tunneling~\cite{Coleman:1977py,Callan:1977pt,Steinhardt:1981ct} or thermal fluctuation~\cite{Linde:1981zj}. 
After nucleation, bubbles expand rapidly~\cite{Cai:2020djd,Wang:2022txy,Lewicki:2022nba,Wang:2023kux,Wang:2023lam,Yuwen:2024hme,Krajewski:2024gma} and eventually percolate with violent collisions, during which stochastic gravitational-wave backgrounds (SGWBs) are generated via bubble collisions~\cite{Witten:1984rs,Kosowsky:1991ua,Kosowsky:1992vn,Kamionkowski:1993fg,Huber:2008hg,Cutting:2018tjt,Cutting:2020nla,Lewicki:2020jiv,Ellis:2020nnr,Lewicki:2020azd,Lewicki:2022pdb,Wang:2025eee}, sound waves~\cite{Hogan:1986qda, Hindmarsh:2013xza, Hindmarsh:2015qta, Hindmarsh:2017gnf, Jinno:2017fby}, and magnetohydrodynamic (MHD) turbulence~\cite{Kamionkowski:1993fg, Kosowsky:2001xp, Dolgov:2002ra, RoperPol:2022iel}. The resulting SGWBs are potentially detectable by currently operating interferometers, such as LIGO–Virgo~\cite{Romero:2021kby,Huang:2021rrk,Jiang:2022mzt,Badger:2022nwo,Yu:2022xdw} and pulsar timing arrays (PTAs)~\cite{Xu:2023wog,NANOGrav:2023gor,NANOGrav:2023hvm,EPTA:2023sfo,Reardon:2023gzh}, as well as by future missions such as LISA~\cite{LISA:2017pwj,Baker:2019nia}, Taiji~\cite{Hu:2017mde,Ruan:2018tsw}, and TianQin~\cite{TianQin:2020hid,TianQin:2015yph}.
Furthermore, FOPTs may induce other characteristic phenomena such as curvature perturbations~\cite{Liu:2022lvz}, baryogenesis (or leptogenesis)~\cite{Kuzmin:1985mm, Trodden:1998ym,Morrissey:2012db,Huang:2022vkf,Ai:2025vfi}, primordial black holes (PBHs)~\cite{Liu:2021svg,Hashino:2021qoq,He:2022amv,Kanemura:2024pae,Hashino:2025fse,Franciolini:2025ztf,Ning:2026nfs}, or even a baby universe~\cite{Sato:1981gv, Cao:2025jwb}.

In quantum chromodynamics (QCD), chiral symmetry is a fundamental symmetry that explains quark condensation and the masses of hadrons~\cite{Ecker:1994gg}. 
It has been predicted that the chiral phase transition in QCD can be first-order if the flavor number $N_{f}$ of massless fermions satisfies $N_{f} \geq 3$~\cite{Pisarski:1983ms}. 
This expectation has been numerically investigated using the lattice simulation~\cite{Brown:1990ev, Philipsen:2021qji}. 
Beyond QCD, chiral symmetry plays a crucial role in a variety of dark matter (DM) models, including strongly interacting massive particle (SIMP) dark matter~\cite{Hochberg:2014dra,Hochberg:2014kqa} and composite DM scenarios~\cite{Davoudiasl:2012uw,Bai:2013xga,LatticeStrongDynamicsLSD:2013elk, Tsumura:2017knk}. 
In this study, we investigate the chiral phase transition in the early universe using the three-flavor Nambu–Jona-Lasinio (NJL) model~\cite{Nambu:1961tp,Nambu:1961fr}, which has been proven to be an effective phenomenological framework for describing both chiral symmetry breaking and its restoration at finite temperature and density (e.g., see Ref.~\cite{Klevansky:1992qe,Hatsuda:1994pi}).
The chiral phase transition has been actively investigated over several decades~\cite{Roberts:1994dr,Alkofer:2000wg,Fischer:2006ub,Schwaller:2015tja,Helmboldt:2019pan,YuanyuanWang:2022nds,Lu:2025pqz,Mei:2025ein,Wan:2025wdg,Jiang:2025ofd,Kang:2025nhe,Hua:2026lcm}, and its dynamics in curved spacetime has also been analyzed in various contexts~\cite{Inagaki:1994ec,Elizalde:1994zv,Kanemura:1994rs,Kanemura:1995sx, Inagaki:1995bk,Gitman:1996mk,Ishikawa:1996yx,Inagaki:1997kz, Kim:1997ak,Vitale:1998wm,Goyal:2000yx,Flachi:2010yz,Flachi:2011sx,Flachi:2014jra,Flachi:2015fna,DeMott:2022qux}. 
While our primary focus is on the chiral phase transition, we do not restrict the energy scales and parameters of the NJL model to the QCD regime. Instead, we adopt a more general approach, ensuring that our conclusions remain applicable to dark-sector physics across a broad range of dynamical scales, from MeV to TeV, as explored in Refs.~\cite{Schwaller:2015tja,Helmboldt:2019pan,Hua:2026lcm}.

On the other hand, the impact of black holes on the stability of the electroweak vacuum~\cite{Hiscock:1987hn,Berezin:1990qs,Gregory:2013hja,Burda:2015yfa,Mukaida:2017bgd,Kohri:2017ybt,Oshita:2019jan,Gregory:2020cvy,Hayashi:2020ocn,Strumia:2022jil,Rossi:2025fix} and on generic phase transitions~\cite{Quinta:2019hrf,El-Menoufi:2020ron,Shkerin:2021zbf,Shkerin:2021rhy,Briaud:2022few,Jinno:2023vnr,Zhong:2025xwm} has been studied. 
In particular, corrections to the tunneling rate induced by light BHs have been extensively discussed~\cite{Quinta:2019hrf,El-Menoufi:2020ron,Shkerin:2021zbf,Shkerin:2021rhy,Briaud:2022few,Jinno:2023vnr,Zhong:2025xwm}, and it is expected that the tunneling rate can be enhanced in the vicinity of the event horizon through a gravitational catalytic effect.
Since PBHs are also well-motivated candidates for dark matter~\cite{Hawking:1971ei,Carr:1974nx,Carr:2016drx,Sasaki:2016jop}, understanding their influence on the chiral phase transition is of considerable importance. 
However, several previous studies have shown that chiral symmetry, which is spontaneously broken far from the black hole, can be restored near the event horizon~\cite{Flachi:2011sx, Flachi:2015fna, DeMott:2022qux}. 
These results reveal a nontrivial interplay between black hole–induced chiral symmetry breaking and local restoration of chiral symmetry around the event horizon.
A systematic investigation of this interplay in chiral symmetry has not yet been carried out. 
Such an analysis is expected to be essential for understanding the dynamics of chiral phase transitions in the universe with PBHs.

In this work, we investigate the impact of PBHs on the chiral phase transition within the NJL model in a Schwarzschild background. We analyze the resulting phase structure and the associated SGWB signal. On the formal side, we derive the finite-temperature effective potential in curved spacetime and present a compact and physically transparent decomposition of the PBH-modified potential into flat-spacetime and curvature-induced components.
By choosing benchmark parameters such that the system undergoes a first-order chiral phase transition in flat spacetime, we demonstrate that PBH-induced curvature gives rise to a qualitatively new local phase structure in the near-horizon region. 
As the temperature decreases, the system undergoes a sequence of symmetry-breaking transitions: it first experiences a second-order transition, followed by a first-order transition, and finally restores chiral symmetry near the event horizon. We then solve the bounce equation in the PBH background and, for the first time, determine the inverse duration parameter $\beta / H$ as a function of the PBH fraction $f_{\rm PBH}$, thereby quantifying how the abundance of PBHs modifies the timescale of the chiral phase transition.
Finally, we compute the SGWB spectrum and show that the catalyzed chiral phase transition produces signals with a higher peak frequency and a smaller amplitude compared to the flat-spacetime case. As examples, we show the predicted SGWB signals within the milli-Hz band accessible to LISA-like interferometers~\cite{LIGOScientific:2019vic,Schmitz:2020syl} or within the nano-Hz band probed by PTAs such as NANOGrav~\cite{NANOGrav:2023hvm}.

The rest of the paper is organized as follows. In section~\ref{sec:NJL in curved spacetime}, we construct the thermal effective potential in the NJL model in curved spacetime. In section~\ref{sec:PBH-modified chiral phase transition}, we apply the framework to a Schwarzschild metric and reveal the importance of the curvature-induced contributions to the chiral phase transition. In section~\ref{sec:GW}, we compute the bounce solutions, extract the phase transition parameters $\beta/H$ and $\alpha$, and quantify the resulting GW spectra. Our conclusions and discussions are given in section~\ref{sec:discussions} and section~\ref{sec:conclusions}, respectively.

\section{NJL model in curved spacetime}
\label{sec:NJL in curved spacetime}

In this section, we derive the effective potential that governs the chiral symmetry breaking in curved spacetime.
Throughout the paper, we adopt the metric signature $(-,+,+,+)$ and the natural units with $\hbar=c=k_B=1$.

\subsection{General formalism of the three-flavor NJL model}

In the present work, we focus on strongly interacting systems, which is assumed to be described by a QCD-type theory. In particular, the chiral phase transition is expected to be driven by non-perturbative dynamics, thereby requiring a theoretical treatment beyond standard perturbation theory. To capture the essential features of dynamical chiral symmetry breaking, we employ the three-flavor NJL model as an effective low-energy description. The three-flavor NJL model is characterized by the approximate global symmetry $SU(3)_V \otimes SU(3)_A \otimes U(1)_V$~\cite{Helmboldt:2019pan}, with the action given by
\begin{align}
   \mathcal{S}=\int\md^4 x \sqrt{-g}\left\{ \operatorname{Tr} \bar{\psi}\left(i \slashed{\nabla}-m_0\right) \psi+2 G_S \operatorname{Tr} \Phi^{\dagger} \Phi+G_{A}(\operatorname{det} \Phi+\det \Phi^\dagger)\right\},\label{2-NJL}
\end{align}
where $m_0$ denotes the current mass, and $\Phi$ is a composite field constructed from the fermion field $\psi$, defined as
\begin{align}
\begin{alignedat}{2}
\Phi_{ij} & \equiv \bar{\psi}_{i}(1-\gamma_{5})\psi_{j} 
& = \tfrac{1}{2}\lambda^{a}_{ji}\,\mathrm{Tr}\,\bar{\psi}\lambda^{a}(1-\gamma_{5})\psi \,, \\
(\Phi^\dagger)_{ij} & \equiv \bar{\psi}_{i}(1+\gamma_{5})\psi_{j} 
& = \tfrac{1}{2}\lambda^{a}_{ji}\,\mathrm{Tr}\,\bar{\psi}\lambda^{a}(1+\gamma_{5})\psi \,, 
\end{alignedat}
\end{align}
here $\lambda^a$ are the Gell-Mann matrices with 
$\lambda^0=\sqrt{2/3} \mathbf{1}$. 
We note that $\nabla_\mu$ and $g$ denote the spinorial covariant derivative and the determinant of the spacetime metric $g_{\mu\nu}$, respectively.
For a Dirac spinor $\psi$, the covariant derivative is defined as~\cite{Parker:2009uva}
\begin{align}
    \nabla_\mu \psi = \left(\partial_\mu + \Omega_\mu\right)\psi,
    \quad
    \Omega_\mu = \frac{1}{8}\omega_{\mu}{}^{AB}[\overline{\gamma}_A,\overline{\gamma}_B],\quad \omega_{\mu}{}^{AB}=e_{\nu}{}^{A}\Gamma^\nu{}_{\rho\mu}e^{\rho B}+e_{\nu}{}^{A}\partial_{\mu}e^{\nu B}
\end{align}
where $\Omega_\mu$ is the spin connection and $\Gamma^\nu{}_{\rho\mu}$ is the Christoffel symbol.
The curved-spacetime $\gamma$ matrices are defined by $\gamma^{\mu}(x)\equiv e^{\mu}{}_{A}(x)\,\overline{\gamma}^{A}$, with $e^{\mu}{}_{A}(x)$ and $\overline{\gamma}^{A}$ being the tetrad and the flat-spacetime $\gamma$ matrices, respectively. The second term in Eq.~\eqref{2-NJL}, often referred to as the symmetric term, preserves the full $U(3)_V \otimes U(3)_A$ symmetry. In contrast, the third term, known as the anomaly term, leaves only the reduced symmetry $SU(3)_V \otimes SU(3)_A \otimes U(1)_V$. In QCD, this term is introduced to account for the $\eta$–$\eta^\prime$ mixing and the unusually large mass of the $\eta^\prime$ meson~\cite{Kobayashi:1970ji,tHooft:1976snw}. 

To carry out explicit calculations within the NJL model, we adopt the self-consistent mean-field (SCMF) approximation~\cite{Kobayashi:1970ji,Hatsuda:1994pi,Helmboldt:2019pan}. In this framework, the expectation value of the fermion bilinear $\langle \Phi \rangle$ is expressed in terms of the effective mesonic degrees of freedom as
\begin{align}
    -4G_S\langle\Phi\rangle
    = (\sigma+i \eta')\mathbf{1}
    + 2\left(a_{a} + i \pi_{a}\right)\lambda^{a} \,, 
    \label{2-VEV}
\end{align}
where each meson field is defined by
\begin{align}
    \begin{aligned}
    \sigma = -\frac{4G_S}{3}\langle\bar{\psi}\psi\rangle,\quad
    \pi_a = -4 i G_S\langle\bar{\psi}\gamma_5\lambda_a\psi\rangle,\\
    \eta^{\prime} = -\frac{4 i G_S}{3}\langle\bar{\psi}\gamma_5\psi\rangle,\quad
    a_a = -4G_S\langle\bar{\psi}\lambda_a\psi\rangle. 
    \end{aligned}\label{2-VEV-Meson}
\end{align}
The above mesons field can also be understood as auxiliary fields in the path integral bosonization, which implies that they are non-propagating at tree-level. However, after integrating the dynamical fermion fields out, the kinetic terms of mesons can be induced at the loop-level (See Appendix~\ref{app:meson-propagator}).

A nonzero vacuum expectation value (VEV) of the chiral condensation $\sigma$ serves as the order parameter for spontaneous chiral symmetry breaking. We therefore focus on the effective potential for $\sigma$, obtained by integrating out the fermionic degrees of freedom~\cite{Inagaki:1997kz,Parker:2009uva}:
\begin{align}
    \label{sec2:V_eff_origin}
    V_{\mathrm{eff}}(\sigma)
    = V_{\mathrm{tree}}\!\left(\sigma,\,\eta'=\pi_a=a_a=0\right)
    + i \ln \Det \, \bigl(i\slashed{\nabla} - M(\sigma)\bigr) \,, 
\end{align}
where $\Det$ denotes the functional determinant of the Dirac operator $i\slashed{\nabla}-M(\sigma)$ obtained after integrating out the fermions. 
At tree level, setting $\eta'=\pi_a=a_a=0$ yields~\cite{Helmboldt:2019pan}
\begin{align}
    V_{\mathrm{tree}}\!\left(\sigma,\,\eta'=\pi_a=a_a=0\right)
    = \frac{3}{8G_S}\sigma^2-\frac{G_A}{16G_S^3}\sigma^3.
\end{align}
The field-dependent constituent mass $M(\sigma)$ in eq.~\eqref{sec2:V_eff_origin} is given by
\begin{align}
    M(\sigma)
    = m_{0} + \sigma - \frac{G_A}{8G_S^{\,2}}\,\sigma^{2}.\label{sec2:Constitute mass}
\end{align}
Using the Schwinger proper-time representation~\cite{Haymaker:1986tt}, the second term in Eq.\,\eqref{sec2:V_eff_origin} can be written as
\begin{align}
\begin{aligned}
    \ln\Det\,\bigl(i\slashed{\nabla}-M(\sigma)\bigr)
    &= \Tr \int \md^4 x\, \ln \,\bigl(i\slashed{\nabla}-M(\sigma)\bigr) \\
    &= -\,\Tr \int \md^4 x \sqrt{-g}\int_{M(0)}^{M(\sigma)} \md s\, S(x,x;s) + \mathrm{const},
    \label{sec2:determinant}
\end{aligned}
\end{align}
where ``$\mathrm{const}$" denotes $\sigma$-independent terms, and $S(x,x;s)$ is the spinor two-point function satisfying the Dirac equation
\begin{align}
    \left(i\slashed{\nabla}-s \right) S(x,y;s)
    = \frac{1}{\sqrt{-g}}\delta^{(4)}(x-y).
    \label{sec2:Dirac_Eq}
\end{align}
Combing Eq.~\eqref{sec2:V_eff_origin} with Eq.~\eqref{sec2:determinant}, the effective potential is given by
\begin{align}
\label{sec2:NJL general effective potential}
    V_\mathrm{eff}(\sigma)
    = \frac{3}{8G_S}\sigma^2-\frac{G_A}{16G_S^3}\sigma^3
    - 3i N_c\Tr\!\int_{M(0)}^{M(\sigma)} \md s\, S(x,x;s),
\end{align}
where $3N_c$ come from the trace in the flavor and color space. Once the two-point function $S(x,x;s)$ is determined, the explicit form of the effective potential follows directly. 

\subsection{Effective potential in the NJL model at finite temperature}

After obtaining the two–point function for fermions in curved spacetime, we now turn to its thermal corrections, which are essential for studying the chiral phase transition in the early universe. At finite temperature, the fermionic two–point function $S_T(x,y;s)$ is defined as the canonical ensemble average
\begin{align}
    S_T(x,y;s)
    = \Tr\!\left[ \rho(T)\,\mathcal{T}\bigl(\psi(x)\bar{\psi}(y)\bigr) \right],
\end{align}
where $T$ is the temperature and $\mathcal{T}$ denotes the time ordering. The density matrix $\rho(T)$ of the canonical ensemble is given by
\begin{align}
    \rho(T)
    \equiv \frac{1}{Z(T)}\sum_\alpha e^{- E_\alpha/T}\,|\alpha\rangle\langle\alpha|,
    \qquad
    Z(T) = \sum_\alpha e^{-E_\alpha/T},
\end{align}
where $Z(T)$ represents the partition function and $E_\alpha$ is the energy eigenvalue of the corresponding eigenstate $|\alpha\rangle$.

To extract the finite–temperature contribution to the effective potential, we employ the imaginary–time formalism of thermal field theory~\cite{Quiros:1999jp}. In this approach, the zero–temperature Feynman rules are modified according to
\begin{align}\label{sec2:Feynman rules in FTFT}
\begin{aligned}
    \int\!\frac{\md^4 p}{(2\pi)^4}
    \ \longrightarrow\ 
    i T\sum_{n=-\infty}^{\infty}
    \int\!\frac{\md^3 p}{(2\pi)^3},
\end{aligned}
\end{align}
where $n$ is an integer that labels the fermionic Matsubara frequencies. 

Applying the above replacement to eq.~\eqref{eq:Veff_flat}, the finite temperature potential of the NJL model in flat spacetime is expressed as 
\begin{align}
\label{eq:Vtot_flat}
V_{\rm eff}^{\rm flat}(\sigma;T) = V_{00}(\sigma) + V_{0T}(\sigma; T) \,, 
\end{align}
where the first term denotes the contribution from tree and one-loop levels at zero temperature, and the second term corresponds to the thermal correction. 
The explicit form of $V_{00}(\sigma)$ and $V_{0T}(\sigma; T)$ is given by 
\begin{align}
    V_{00}(\sigma)&=\frac{3}{8 G_S}\sigma^2- \frac{G_A}{16 G_S^3}\sigma^3 -\frac{N_fN_c\Lambda^4}{16\pi^2}\Bigg[ \ln\left(1+\frac{M(\sigma)^2}{\Lambda^2}\right) \notag \\
    &\qquad\qquad-\left(\frac{M(\sigma)}{\Lambda}\right)^4\ln\left(1+\frac{\Lambda^2}{M(\sigma)^2}\right)+\left(\frac{M(\sigma)}{\Lambda}\right)^2\Bigg] \,, \label{sec3:V00}\\
    V_{0T}(\sigma;T)&=-\frac{2 N_{f} N_{c} T^4}{\pi^2}
    \left[ J_F\left(\frac{M(\sigma)}{T}\right)-J_F\left(\frac{M(0)}{T}\right) \right] \,, \label{sec3:V0T}
\end{align}
where $\Lambda$ denotes the cutoff scale, and the thermal fermionic function $J_F(y)$ is defined as~\cite{Quiros:1999jp}
\begin{align}\label{sec3:thermal fermionic function}
    J_F(y)=\int_0^\infty dxx^2\ln\left[1+e^{-\sqrt{x^2+y^2}}\right].
\end{align}

\begin{figure}[t]
    \centering
    \includegraphics[width=0.6\textwidth]{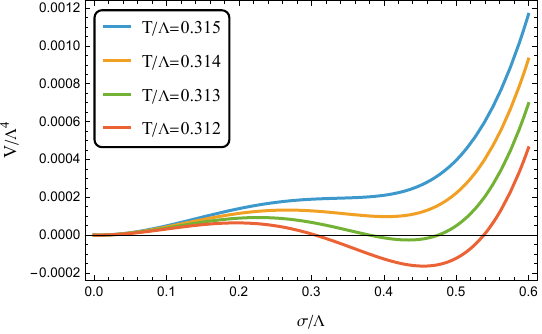}
    \caption{
        Temperature dependence of the flat-spacetime effective potential
        $V_{\rm eff}^{\rm flat}(\sigma, T)$ for the parameter choice
        $G_S\Lambda^2 = 8$, $G_A \Lambda^6 = -448$ and
        $m_0\Lambda^{-1} = 10^{-3}$, illustrating a first-order chiral phase transition.
    }
    \label{Fig:VT}
\end{figure}

The temperature dependence of the effective potential in eq.~\eqref{eq:Vtot_flat} is shown in Fig.~\ref{Fig:VT}, where the parameters are taken as follows:
\begin{align}
    G_S \Lambda^2 = 8 \,, \quad
    G_A \Lambda^6 = -448 \,,  \quad
    m_0 \Lambda^{-1} = 10^{-3} \,. 
    \label{sec3:BenchMark}
\end{align}
As illustrated in Fig.~\ref{Fig:VT}, the effective potential in eq.~\eqref{eq:Vtot_flat} indeed exhibits a first-order chiral phase transition.
At high temperature, thermal effects restore chiral symmetry and the global minimum is located at $\langle \sigma \rangle = 0$. As the temperature decreases, a second local minimum develops at a nonzero value $\langle \sigma \rangle \neq 0$, which eventually becomes the true vacuum, while the symmetric minimum at $\sigma = 0$ remains as a metastable false vacuum. The potential barrier separating these two minima prevents a smooth crossover and instead allows the system to undergo a first-order transition via bubble nucleation. The resulting GW spectrum will be discussed in Sec.~\ref{sec:GW}.

By applying the replacement rules in eq.~\eqref{sec2:Feynman rules in FTFT} to the curved spacetime effective potential in eq.~\eqref{eq:Veff_flat}, we obtain the finite temperature NJL effective potential in a general curved spacetime background.
In the next section, we apply this general finite–temperature formalism to the Schwarzschild geometry surrounding a PBH, and analyze how the gravitational effects alter the dynamics of the chiral phase transition.

\subsection{NJL effective potential in curved spacetime}

As shown in eq.~\eqref{sec2:NJL general effective potential}, the evaluation of the effective potential reduces to determining the two–point function $S(x,x;s)$. To compute it in a curved spacetime background, it is convenient to employ Riemann normal coordinates centered at a spacetime point $x_0$, where the metric and Christoffel symbol take the locally flat form
\begin{align}
g_{\mu\nu}(x_0)  = \eta_{\mu\nu} \,, \quad \Gamma^\alpha{}_{\mu\nu}(x_0)  = 0 \,, 
\end{align}
where $\eta_{\mu\nu}$ is the metric in Minkowski spacetime. In a neighborhood of $x_0$, the metric can be expanded in terms of spacetime displacement $\Delta x \equiv x-x_0$ as~\cite{Parker:2009uva}
\begin{align}
\begin{aligned}
\label{sec2:Metric expansion}
g_{\alpha\beta}(x)
&=\eta_{\alpha\beta}
-\frac{1}{3}R_{\alpha\mu\beta\lambda}(x_0)\,\Delta x^\mu\Delta x^\lambda
-\frac{1}{3!}R_{\alpha\gamma\beta\lambda;\mu}(x_0)\,\Delta x^\lambda\Delta x^\mu\Delta x^\gamma  \notag\\
&\quad +\frac{1}{5!}\!\left[
-6R_{\alpha\delta\beta\gamma;\lambda\mu}(x_0)
+\frac{16}{3}R_{\lambda\beta\mu}{}^{\rho}
   R_{\gamma\alpha\delta\rho}(x_0)
\right]
\Delta x^\lambda\Delta x^\mu\Delta x^\gamma\Delta x^\delta
+\cdots,
\end{aligned}
\end{align}
where $R_{\alpha\beta\mu\nu}$ and $R_{\alpha\beta\mu\nu;\lambda}$ are the Riemann tensor and its covariant derivative, respectively. To simplify the calculation, one can introduce the bispinor function defined by
\begin{align}
    S(x_0,x;s) \equiv \left(i\gamma_\mu\nabla^\mu + s\right)\mathcal{G}(x_0,x;s) \,. 
\end{align}
Based on the property eq.~\eqref{sec2:Dirac_Eq}, it can be proved that $\mathcal{G}(x_0,x;s)$ satisfies~\cite{Parker:2009uva}
\begin{align}
    \left(g^{\mu\nu}\nabla_\mu\nabla_\nu - \frac{1}{4}R + s^{2}\right)\mathcal{G}(x_0,x;s)
    = -\frac{1}{\sqrt{-g}}\delta^{(4)}(x-x_0)\,\mathbf{1} \,, 
\end{align}
where the covariant derivatives act on $\mathcal{G}(x_0,x)$ as a spinor at $x_0$ and $R=g^{\mu\nu}R^{\alpha}{}_{\mu\alpha\nu}$ is the Ricci scalar by definition.
Implementing a local momentum-space expansion in Riemann normal coordinates~\cite{Bunch:1979uk}, we can get 
\begin{align}\label{sec2:Gdefine}
\mathcal{G}(x_0,x;s)
= |g(x)|^{-1/4}\!\int\!\frac{\md^4 k}{(2\pi)^4}
e^{ik\cdot(x-x_0)}\,\mathcal{G}(k;s) \,, 
\end{align}
with
\begin{align}
\mathcal{G}(k;s)
&=
\Biggl[
1
+ \left(A_1 + i A_{1\alpha}\frac{\partial}{\partial k_\alpha}
      -A_{1\alpha\beta}\frac{\partial}{\partial k_\alpha}
                           \frac{\partial}{\partial k_\beta}
   \right)\!\left(-\frac{\partial}{\partial s^{2}}\right)
+ A_2 \left(\frac{\partial}{\partial s^{2}}\right)^2
\Biggr]
(k^2-s^2)^{-1},\label{sec2:G in k space}
\end{align}
where the coefficients are given by~\cite{Parker:2009uva}
\begin{align}
A_{1}   &= \frac{1}{12} R \mathbf{1} \,,  \label{sec2:Coefficient A1}\\
A_{2}   &= \left(+\frac{1}{120} R_{; \mu}{ }^{\mu}+\frac{1}{288} R^{2}-\frac{1}{180} R_{\mu \nu} R^{\mu \nu}+\frac{1}{180} R_{\mu \nu \sigma \tau} R^{\mu \nu \sigma \tau}\right) \\ \notag
&\quad+\frac{1}{48} \Sigma_{[\alpha \beta]} \Sigma_{[\gamma \delta]} R^{\alpha \beta \lambda \chi} R^{\gamma \delta}{ }_{\lambda \chi} \,, \label{sec2:Coefficient A2} \\
A_{1 \mu}   &= -\frac{1}{24} R_{; \mu} 1-\frac{1}{12} \Sigma_{[\alpha \beta]} R^{\alpha \beta \lambda}{ }_{\mu ; \lambda} \,, \\
A_{1 \mu \nu}   &= \frac{1}{30}\left(\frac{1}{4} R_{\mu \nu ; \lambda}-\frac{1}{2} R_{; \mu \nu}+\frac{1}{3} R_{\mu \lambda} R^{\lambda}{ }_{\nu}\right. \left.-\frac{1}{6} R^{\lambda \chi} R_{\lambda \mu \chi \nu}-\frac{1}{6} R^{\lambda \chi \sigma \mu} R_{\lambda \chi \sigma \nu}\right) \mathbf{1}\notag \\
&\quad +\frac{1}{48} \Sigma_{[\alpha \beta]}\left(R R^{\alpha \beta}{ }_{\mu \nu}+R^{\alpha \beta \lambda}{ }_{\mu ; \lambda \nu}+R^{\alpha \beta \lambda}{ }_{\nu ; \lambda \mu}\right)\notag\\
&\quad-\frac{1}{96} \Sigma_{[\alpha \beta]} \Sigma_{[\gamma \delta]}\left(R^{\alpha \beta \lambda}{ }_{\mu} R^{\gamma \delta}{ }_{\lambda \nu}+R^{\alpha \beta \lambda}{ }_{\nu} R^{\gamma \delta}{ }_{\lambda \mu}\right) \,, \label{sec2:Coefficient A1uv}
\end{align}
and 
\begin{align}
    \Sigma_{[\alpha\beta]} \equiv \frac{1}{4}\left[\gamma_\alpha,\gamma_\beta\right].
\end{align}
In the coincidence limit $x\to x_0$, the spinor two–point function becomes
\begin{align}\label{sec2:relation S and G}
    S(x,x;s)
    = \int \frac{\md^4 k}{(2\pi)^4}
    \left(-\gamma^\mu k_\mu + s\right) \mathcal{G}(k;s).
\end{align}
Using eq.~\eqref{sec2:NJL general effective potential}, we finally obtain the effective potential in curved spacetime,
\begin{align}
\label{eq:Veff_flat}
    V_\mathrm{eff}(\sigma)
    = \frac{3}{8 G_S}\sigma^{2}
    - \frac{G_A}{16 G_S^{3}}\sigma^{3}
    - iN_f N_c\Tr\!\int_{M(0)}^{M(\sigma)} \md s
      \int \frac{\md^4 k}{(2\pi)^4}
      \left(-\slashed{k}+s\right)\mathcal{G}(k;s).
\end{align}

\section{PBH-modified chiral phase transition}
\label{sec:PBH-modified chiral phase transition}

In the previous section, we have derived the finite temperature effective potential within the NJL model in a general curved spacetime. We then apply this formalism to a specific gravitational background that is directly relevant to the PBH.

PBHs provide localized regions of strong gravity embedded in an otherwise approximately homogeneous cosmological plasma. Although the Schwarzschild spacetime describing an isolated, non-rotating PBH is Ricci-flat, $R_{\mu\nu}=0$, it is not trivial from the perspective of quantum fields. Higher-order curvature invariants, such as $R_{\mu\nu\sigma\tau}R^{\mu\nu\sigma\tau}$, remain nonzero and contribute to the effective action, in particular via the coefficient $A_2$ in eq.~\eqref{sec2:Coefficient A2}. As a consequence, both spacetime curvature and the local thermal environment near a PBH can significantly modify the pattern of chiral symmetry breaking and its restoration compared to the flat and homogeneous case.

\subsection{PBH-modified effective potential}
\label{sec3.1:potential}

We consider an isolated and non-rotating PBH described by the Schwarzschild metric. Neglecting the cosmological expansion on scales much smaller than the Hubble parameter, the spacetime geometry in the vicinity of the PBH is well approximated by
\begin{align}\label{sec3:Schwarzschild metric}
    \md s^2
    = -f(r)\md t^2
      + f(r)^{-1} \md r^2
      + r^2 \left( \md\theta^2 + \sin^2\theta \,\md\phi^2 \right)  ~~ \text{with} ~~ f(r)= 1 - \frac{r_s}{r},
\end{align}
where $r_s \equiv 2 G_N M_{\PBH}$ is the Schwarzschild radius of a PBH with mass $M_{\PBH}$, and $G_N$ is Newton’s gravitational constant. In what follows, we use this background to evaluate the curvature-dependent terms in the effective potential and to isolate the PBH-induced corrections to the chiral dynamics.

Using the metric~\eqref{sec3:Schwarzschild metric}, the traced coefficients in eqs.~\eqref{sec2:Coefficient A1}–\eqref{sec2:Coefficient A1uv} can be explicitly evaluated as
\begin{align}\label{sec3:Coefficients}
    \Tr A_1 = \Tr A_{1\mu} = 0,
    \quad
    \Tr A_2 = - \Tr A_{1\mu}{}^{\mu}
    = \frac{7 r_s^2}{30 r^6} \,. 
\end{align}
Substituting the coefficients~\eqref{sec3:Coefficients} into the curved-spacetime effective potential~\eqref{eq:Veff_flat} and applying the finite-temperature Feynman rules in Eq.~\eqref{sec2:Feynman rules in FTFT}, the effective potential can be written as the sum of four contributions  (see Appendix~\ref{appA} for details):
\begin{align}
\begin{aligned}
    V_\mathrm{eff}(\sigma;T,r)
    &= \frac{3}{8 G_S}\sigma^2
       - \frac{G_A}{16 G_S^3}\sigma^3
       - 12 N_f  \int_{M(0)}^{M(\sigma)} \md s\, s
         \int \frac{\md^3 k}{(2\pi)^3} \frac{1}{2\omega_k} \\
    &\quad + 12 N_c \int_{M(0)}^{M(\sigma)} \md s
         \int \frac{\md^3 k}{(2\pi)^3}
         \frac{1}{\omega_k} \frac{1}{e^{\omega_k/T} + 1}\\
    &\quad + 4 N_c \int_{M(0)}^{M(\sigma)} \md s\, s^3
         \frac{7 r_s^2}{30 r^6}
         \left(\frac{\partial}{\partial s^2}\right)^3
         \int \frac{\md^3 k}{(2\pi)^3}\frac{1}{2\omega_k} \\
    &\quad - 4 N_c \int_{M(0)}^{M(\sigma)} \md s\, s^3
         \frac{7 r_s^2}{30 r^6}
         \left(\frac{\partial}{\partial s^2}\right)^3
         \int \frac{\md^3 k}{(2\pi)^3}
         \frac{1}{\omega_k}\frac{1}{e^{\omega_k/T} + 1},
         \label{sec3:effective potential-1}
\end{aligned}
\end{align}
where $\omega_k \equiv \sqrt{|\boldsymbol{k}|^{\,2} + s^2}$. 
For later convenience, we denote four parts of $V_\mathrm{eff}(\sigma)$ by
\begin{align}\label{sec3:full potential into 4 pieces}
    V_\mathrm{eff}(\sigma;T,r)
    = V_{\rm eff}^{\rm flat}(\sigma, T)
      + V_{R0}(\sigma;r) + V_{RT}(\sigma;T,r),
\end{align}
where $V_{\rm eff}^{\rm flat}(\sigma, T)$ is given in eq.~\eqref{eq:Veff_flat}. 
$V_{R0}(\sigma;r)$ and $V_{RT}(\sigma;T,r)$ encode the curvature-induced contribution of the PBH background at zero temperature and finite temperature, respectively. 
After some manipulations (see Appendix~\ref{appA}), $V_{R0}(\sigma;r)$ and $V_{RT}(\sigma;T,r)$ are given by
\begin{align}
    V_{R0}(\sigma;r)&=N_fN_c\frac{7r_s^2}{480\pi^2r^6} \ln\left|\frac{M(\sigma)}{m_0}\right| \,, \label{sec3:VR0}\\
    V_{RT}(\sigma;T,r)&=N_f N_c\frac{7r_s^2} {60 r^6} \frac{2 T^4}{\pi^2} \int_{M(0)}^{M(\sigma)}ds s^3 \left( \frac{\partial}{\partial s^2}\right)^4 J_F\left(\frac{s}{T} \right) \,. \label{sec3:VRT}
\end{align}

\subsection{Curvature corrections to the effective potential at the zero temperature \label{subsec:curvanture induced corrections}}

Switching on the PBH background amounts to including the curvature-dependent contributions $V_{R0}(\sigma;r)$ and $V_{RT}(\sigma;T,r)$ in the effective potential.
To most transparently expose the symmetry-restoring tendency of curvature, we first consider the zero-temperature limit.
In this case, the effective potential reduces to
\begin{align}
V_\mathrm{eff}(\sigma;T = 0,r)
 = V_{\rm eff}^{\rm flat}(\sigma, T) + V_{R0}(\sigma;r) \,. 
\end{align}
Since the curvature-induced term $V_{R0}(\sigma;r)$ scales as $r_s^2/r^6$, it becomes relevant only in the near-horizon region $r \sim \mathcal{O}(r_s)$.

\begin{figure}[t]
    \centering
    \includegraphics[width=0.5\textwidth]{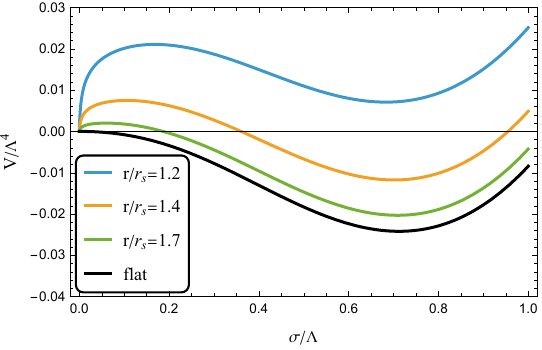}
    \caption{
    Radial dependence of the zero-temperature effective potential
    $V_R(\sigma;r)=V_{00}(\sigma)+V_{R0}(\sigma;r)$ in the Schwarzschild PBH background for the benchmark parameters in eq.~\eqref{sec3:BenchMark}.
    The colored curves show $V_R(\sigma;r)$ at fixed radii $r/r_s=1.2,\,1.4,\,1.7$, while the black curve corresponds to the flat-spacetime potential $V_{00}(\sigma)$.
    The PBH-induced curvature term lifts the broken-phase minimum and thus tends to restore chiral symmetry as $r\to r_s$.
    }
    \label{Fig:VR}
\end{figure}

In figure~\ref{Fig:VR}, the radial direction dependence of the effective potential at zero temperature $V_{\rm eff}(\sigma; T = 0, r)$ is shown. 
This figure indicates that the PBH contribution raises the free energy of the symmetry-broken phase relative to that of the symmetric phase, thereby favoring chiral symmetry restoration.
This restoration effect becomes increasingly pronounced as one approaches the horizon because of the rapid growth of the $r_s^2/r^6$ factor in eq.~\eqref{sec3:VR0}.
The barrier separating the symmetric minimum at $\sigma=0$ and the broken minimum at $\sigma\neq 0$ is essential for the curvature-induced symmetry restoration discussed in Sec.~\ref{subsec:local phase structure}.
Moreover, this finding motivates the notion of a localized symmetry-restored shell surrounding the PBH horizon and its possible implications for Hawking emission. 
On the other hand, at sufficiently large radii, the gravitational contribution $V_{R0}(\sigma;r)$ in eq.~\eqref{sec3:VR0} is strongly suppressed and the potential approaches its flat spacetime form, so the symmetry remains broken far from the PBH.
This behavior is consistent with earlier analyses of dynamical symmetry breaking in the Schwarzschild background~\cite{Flachi:2011sx}.

\subsection{Finite temperature potential around a PBH \label{subsec:local phase structure}}

We then discuss the finite temperature potential around a PBH and show that a non-trivial local phase structure appears around the event horizon. 

As shown in Fig.~\ref{Fig:VR}, due to the curvature effect, the symmetry restoration can occur as $r \to r_{s}$.
On the other hand, thermal corrections also achieve the symmetry restoration at high temperatures~\cite{Linde:1981zj}. 
The combination of these independent effects gives rise to a non-trivial phase structure around PBHs.
Fixing the radial position very close to the PBH, $r/r_s = 1.05$, we trace how the local potential $V_\mathrm{eff}(\sigma;T,r)$ evolves as the cosmological temperature decreases.
The resulting behavior is shown in Fig.~\ref{Fig:Vfullrs1r105} and exhibits two qualitatively distinct stages:
\begin{figure}[t]
    \centering
    \begin{subfigure}{0.48\textwidth}
        \centering
        \includegraphics[width=\linewidth]{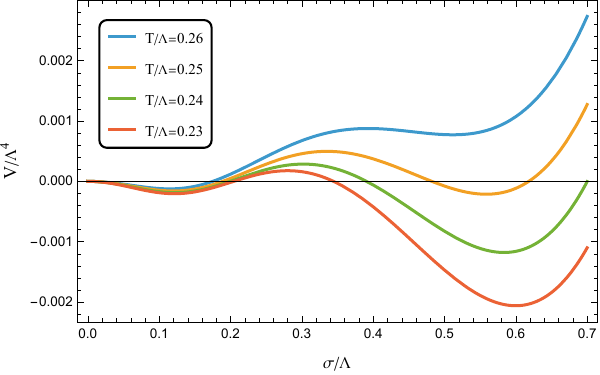}
        \caption{Symmetry-broken process}
        \label{Fig:Vfullrs1r105-b}
    \end{subfigure}
    \begin{subfigure}{0.48\textwidth}
        \centering
        \includegraphics[width=\linewidth]{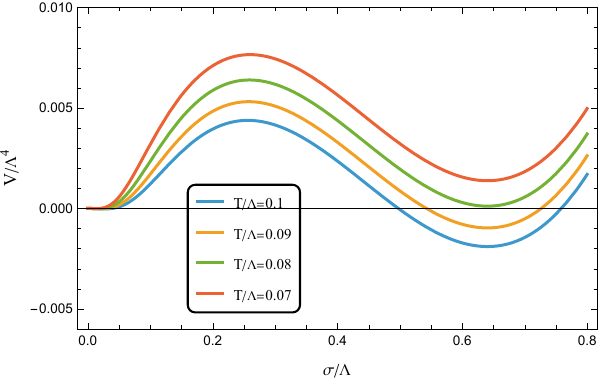}
        \caption{Symmetry-restored process}
        \label{Fig:Vfullrs1r105-c}
    \end{subfigure}
    \caption{
        Temperature dependence of the full effective potential
        $V_\mathrm{eff}(\sigma;T,r)$ at $r/r_s = 1.05$ with $r_s\Lambda = 1$ and the parameters in eq.~\eqref{sec3:BenchMark}.
        Panels (a) and (b) respectively show the FOPT subsequent to the second-order phase transition and the symmetry restoration process.
    }
    \label{Fig:Vfullrs1r105}
\end{figure}

\begin{itemize}

\item \textbf{(a) First-order phase transition following the second-order phase transition:} 
At relatively high temperatures (Fig.~\ref{Fig:Vfullrs1r105-b}), the effective potential has a single minimum at $\sigma = 0$, corresponding to a chiral symmetric phase. 
As the universe cools, the vacuum expectation value of the $\sigma$ field gets a nonzero value ($\sigma/\Lambda \sim 1.4$) continuously without a barrier. 
This behavior characterizes a second-order chiral phase transition at $r/r_s = 1.05$.
As the temperature drops further, the potential gets two degenerate minima separated by a barrier (orange line in Fig.~\ref{Fig:Vfullrs1r105-b}). 
The presence of this potential barrier signals a first-order phase transition. 

\item \textbf{(b) Symmetry restoration due to the curvature effect:}
At even lower temperatures (Fig.~\ref{Fig:Vfullrs1r105-c}), the potential at the origin $\sigma = 0$ becomes the global minimum due to the curvature effect.
This implies that the PBH has the vacuum energy and a non-trivial local phase structure around its event horizon. 
This symmetry restoration at low temperature is a novel feature induced by the PBH curvature corrections and does not occur in the NJL model within the flat spacetime. 

\end{itemize}

The analysis in this section has established how the PBH background reshapes the local free-energy landscape as a function of temperature $T$ and the radial point $r$. 
In the next section, we discuss the phase transition dynamics and calculate finite-temperature bounce solutions with position-dependent potential $V_\mathrm{eff}(\sigma;T,r)$, 
Using the results, we quantify the bubble nucleation rate and the impacts of PBH catalysis effects on phase transition properties such as the duration of the phase transition $\beta/H$.

\section{PBH-catalyzed phase transition and gravitational-wave signals}\label{sec:GW}

In Sec.~\ref{sec:PBH-modified chiral phase transition}, we have shown that the Schwarzschild PBH background induces nontrivial, position-dependent modifications of the chiral effective potential. We now turn to the dynamical consequences of these modifications for the chiral phase transition. In this section, we numerically compute the bounce solutions in the PBH background and study how PBH-induced catalysis of the phase transition affects the resulting stochastic gravitational-wave signal.

\subsection{PBH–catalyzed phase transition \label{subsec:PBH-catalysis}}

To quantify the PBH–catalyzed phase transition, we compute the bounce solutions in the presence of the position–dependent effective potential $V_\mathrm{eff}(\sigma;T,r)$. Throughout this analysis, the PBH background is treated as a fixed classical geometry and the backreaction of the scalar $\sigma$ field on the spacetime metric is neglected.

The meson fields introduced in eq.~\eqref{2-VEV-Meson} are bosonic auxiliary fields. Thus, they do not possess tree-level kinetic terms and are classically non-propagating in the NJL model. Since tunneling processes are inherently dynamical, a description of $\sigma$-field tunneling requires the kinetic term induced by quantum fluctuations. Within this framework, the auxiliary field $\sigma$ is promoted to a propagating quantum field. The key ingredient is the finite-temperature wave-function renormalization,
\begin{align}
    Z_{\sigma}^{-1}(\sigma)
    = \left.\frac{\md \Gamma_{\sigma\sigma}\!\left(p^{2},\sigma\right)}{\md p^{2}}\right|_{p^{2}=0},
    \label{sec3:wave-function Renormalization}
\end{align}
where $\Gamma_{\sigma\sigma}(p^2,\sigma)$ denotes the inverse one-loop propagator of the $\sigma$ field (cf.\,eq.~\eqref{app:inverse propagator}). In our analysis, we employ the wave function renormalization computed in flat spacetime~\cite{Helmboldt:2019pan}, as the gravitational corrections to $Z_{\sigma}$ are suppressed by the loop factor and the curvature coupling ${\rm Tr}A_{2}/\Lambda^4 \ll 1$.

For an $O(3)$-symmetric bounce configuration at finite temperature, the Euclidean action in the Schwarzschild spacetime is given by~\cite{Tong:2023krn,Helmboldt:2019pan}
\begin{align}
    S_3(T)
    = 4\pi \int_{r_s}^{\infty} \md r\, r^2
    \left[
        \frac{Z_\sigma^{-1}(\sigma)\,f(r)}{2}\,\bigl(\sigma'(r)\bigr)^2
        + V_\mathrm{eff}(\sigma;T,r)
    \right],
    \label{sec3:S3-definition}
\end{align}
where the prime denotes the derivative with respect to the radial coordinate $r$. The renormalization factor $Z_\sigma(\sigma)$ is defined in eq.~\eqref{sec3:wave-function Renormalization}, and $f(r)$ is the Schwarzschild lapse function introduced in eq.~\eqref{sec3:Schwarzschild metric}.
The corresponding equation of motion for the bounce profile $\sigma(r)$ is
\begin{align}
    f(r)\,\sigma''(r)
    + \left(\frac{2 f(r)}{r} + f'(r)\right)\sigma'(r)
    + \frac{f(r)}{2}\,\frac{\partial \ln Z_\sigma^{-1}}{\partial\sigma}\,
      \bigl(\sigma'(r)\bigr)^2
    = Z_\sigma(\sigma)\,\frac{\partial V_\mathrm{eff}(\sigma;T,r)}{\partial \sigma} \,. 
    \label{sec4:bounce}
\end{align}
We impose the following boundary conditions:
\begin{align}
\sigma(r \to \infty) = \sigma_f \,, \quad
    \sigma'(r_s)
    = r_s Z_{\sigma}(\sigma)\,\frac{\partial V_\mathrm{eff}(\sigma;T,r)}{\partial \sigma}
      \bigg|_{\sigma=\sigma(r_s)} \,.
    \label{sec4:bounce-BC}
\end{align}
where $\sigma_f$ denotes the VEV of the $\sigma$ field at the false vacuum. As shown in Ref.~\cite{Hayashi:2020ocn}, the boundary condition at the event horizon ensures the regularity of the solution and resolves the coordinate singularity at $r = r_s$ in eq.~\eqref{sec4:bounce}.

In practice, eq.~\eqref{sec4:bounce} is solved numerically using a shooting method. For a fixed temperature $T$, the initial value $\sigma(r_s)$ is adjusted iteratively until the solution converges to the false vacuum $\sigma_f$ at large $r$ within the prescribed tolerance. The resulting bounce profile $\sigma(r)$ is then substituted into eq.~\eqref{sec3:S3-definition} to obtain the three-dimensional Euclidean action $S_3(T)$.

The thermal bubble nucleation rate $\Gamma_0(T)$, interpreted as the decay rate of the false vacuum per unit volume in flat spacetime, is given by~\cite{Linde:1981zj,Witten:1984rs,Hogan:1983ixn}:
\begin{align}
    \Gamma_0(T)
    \simeq T^4 \left(\frac{S_{3}^{\rm flat}(T)}{2\pi T}\right)^{3/2}
    \exp\!\left[-\frac{S_{3}^{\rm flat}(T)}{T}\right],
\end{align}
where \(S_{3}^{\rm flat}(T)\) denotes the three-dimensional Euclidean action evaluated according to eq.~\eqref{sec3:S3-definition} in the flat spacetime limit \(r_s=0\). In the presence of PBHs, the local decay rate is modified. Following Refs.~\cite{Burda:2016mou,Cai:2017tmh}, we consider the purely PBH-catalyzed limit, in which all bubbles present at the time of collision are nucleated in the vicinity of PBHs. 
In this case, the decay rate per unit volume can be written as
\begin{align}
    \Gamma_\mathrm{PBH} (T) \sim T^4\exp\left[-\frac{S_3(T)}{T} \right] \,. 
\end{align}
We note that $ \Gamma_\mathrm{PBH}(T)$ does not have the prefactor with $S_{3}(T)/T$ because of the absence of zero modes corresponding to spatial translation. 
We first define the inverse duration parameters for the cases in flat spacetime and purely PBH-catalyzed limit, respectively:
\begin{align}
    \label{sec4:beta_over_H_def}
    \frac{\beta_0}{H} \equiv -\frac{\md \ln \Gamma_0}{\md \ln T}\Bigg|_{T = T_n} \,, \qquad 
    \frac{\beta_\mathrm{PBH}}{H} \equiv -\frac{\md \ln \Gamma_\mathrm{PBH}}{\md \ln T}\Bigg|_{T = T_n} \,.
\end{align}
Here we determine $T_n$ using the simpler criterion $\Gamma_{0/\PBH}(T_n) \simeq H^4(T_n)$, with $H$ being the Hubble parameter.
For each case, we solve the bounce equation \eqref{sec4:bounce} for the $\sigma$ field and compute the inverse duration parameters $\beta_0/H$ and $\beta_{\rm PBH}/H$ from eq.~\eqref{sec4:beta_over_H_def}.

Then, we briefly discuss the influence of PBHs on the inverse duration and its dependence on the mass of PBHs. The larger value of $\beta/H$ corresponds to a rapid phase transition, while the smaller values indicate a slower, more prolonged transition.
\begin{figure}[t]
    \centering
    \includegraphics[width=0.6\textwidth]{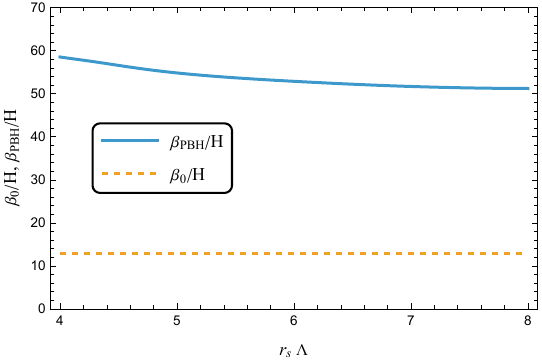}
    \caption{
        Ratio $\beta/H$ as a function of the Schwarzschild radius $r_s$ (in units of $\Lambda^{-1}$) for the benchmark
        $G_S \Lambda^2 = 8$, $G_A \Lambda^5 = -448$, $m_0 \Lambda^{-1} = 10^{-3}$.
        The orange line shows the flat–spacetime result $\beta_0/H$, while the blue curve corresponds to $\beta_{\rm PBH}/H$.
        The presence of a PBH enhances $\beta/H$ and hence catalyzes the phase transition, with the enhancement being larger for lighter PBHs (smaller $r_s$).
    }
    \label{Fig:betaH}
\end{figure}
Figure~\ref{Fig:betaH} shows the resulting ratio $\beta/H$ as a function of the PBH mass, for the benchmark eq.~\eqref{sec3:BenchMark}.
The orange line corresponds to the flat spacetime result and provides a reference value $\beta_0/H$. The blue curve displays the result at the purely PBH-catalysis limits. Two robust features can be identified:
\begin{itemize}
    \item[(i)] For all PBH masses considered, the value of $\beta/H$ in the PBH background is significantly larger than that in the flat spacetime. This implies that the chiral phase transition completes within a much shorter time when PBHs are present. In this sense, the PBH acts as a catalyst that accelerates the transition.

    \item[(ii)] The PBH-induced enhancement of $\beta_{\rm PBH}/H$ decreases mildly as $r_s$ increases. Lighter PBHs, which generate stronger curvature effects near the event horizon, correspond to larger values of $\beta_{\rm PBH}/H$. This behavior is consistent with the scaling $r_s^2/r^6$ for the curvature-induced terms in the effective potential discussed in Sec.~\ref{subsec:local phase structure}.
\end{itemize}

More generally, bubbles can be nucleated in either a homogeneous background far from PBHs or the vicinity of PBHs. 
To account for the combined contributions of homogeneous nucleation and PBH-catalyzed nucleation, we introduce an effective averaged decay rate \(\Gamma(T)\), defined as a weighted sum:
\begin{align}
    \Gamma(T)\equiv\left(1-\frac{N_\mathrm{PBH}}{N_{\mathrm{tot}}}\right)\Gamma_0(T)+\frac{N_\mathrm{PBH}}{N_{\mathrm{tot}}}\Gamma_\mathrm{PBH}(T),\label{sec4:averaged gamma}
\end{align}
where $N_\mathrm{PBH}$ is the number of bubbles nucleated by PBHs within a Hubble volume at the time of bubble collision and $N_\mathrm{tot}$ is the total number of bubbles. 
Assuming that each PBH acts as a nucleation site, the number of bubbles nucleated by PBHs within a Hubble volume is given by
\begin{align}
    N_\mathrm{PBH} 
    = \frac{f_\mathrm{PBH}\,\rho_\mathrm{DM}(T)}{M_\mathrm{PBH}} \cdot \frac{4\pi}{3H^3}
    = \frac{8\pi G_N f_\mathrm{PBH}\,\rho_\mathrm{DM}(T)}{3 r_s H^3},
    \label{sec4:NPBH}
\end{align}
where we have expressed the PBH mass in terms of the Schwarzschild radius, $M_\mathrm{PBH} = r_s / (2G_N)$. The dark matter energy density $\rho_\mathrm{DM}(T)$ is extrapolated from the present day assuming adiabatic expansion:
\begin{align}
    \rho_\mathrm{DM}(T) = \rho_\mathrm{DM}(T_0) \left( \frac{T}{T_0} \right)^3 \frac{g_{*S}(T)}{g_{*S}(T_0)}.
    \label{sec4:rhoDM}
\end{align}
For the numerical evaluation, we adopt the present-day values $T_0\simeq 2.72\,\mathrm{K}$~\cite{Fixsen:2009ug}, $\rho_\mathrm{DM}(T_0)\simeq 1.2\times10^{-6}\,\mathrm{GeV\,cm^{-3}}$~\cite{Planck:2018vyg}, and $g_{*S}(T_0)\simeq 3.9$~\cite{Saikawa:2018rcs}.
In addition, the total number of bubbles at the collision time is approximated by
\begin{align}
    N_\mathrm{tot}=\frac{1}{H^3 R_*^3},\label{sec4:Ntot}
\end{align}
where $R_* \simeq (8\pi)^{1/3} v_w / \beta$ is the mean bubble separation at collision~\cite{Cutting:2020nla,Wang:2025eee}.

\begin{figure}[t]
    \centering
    \includegraphics[width=0.6\textwidth]{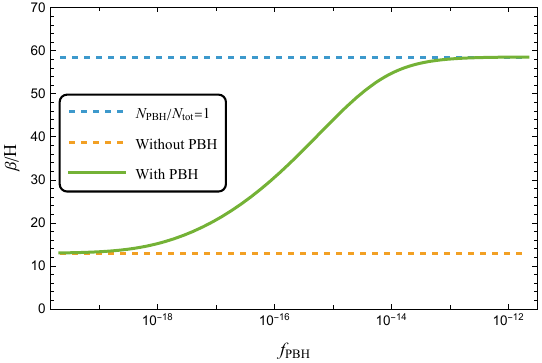}
    \caption{
    Dependence of the phase transition duration $\beta/H$ on the PBH fraction $f_\mathrm{PBH}$ for the benchmark point in Eq.~\eqref{sec3:BenchMark} with $r_{s} \Lambda = 4$ and $\Lambda=1 \,\mathrm{GeV}$.
    The green (orange) line represents the prediction for $\beta/H$ with (without) PBHs.
    The blue line denotes the asymptotic value $\beta_\mathrm{PBH}/H$ in the limit $N_\mathrm{PBH}/N_\mathrm{tot} \to 1$. 
    }
    \label{Fig:betaH_fPBH}
\end{figure}

The parameter $\beta$ quantifies the inverse duration of the phase transition and is defined as:
\begin{align}
       \frac{\beta}{H}
    = -\left. \,
        \frac{\md \ln\Gamma(T)}{\md \ln T}
      \right|_{T = T_n} \,. 
    \label{sec4:beta-over-H}
\end{align}
Substituting eq.~\eqref{sec4:averaged gamma} into the definition of $\beta/H$, and assuming that the ratio $N_\mathrm{PBH}/N_\mathrm{tot}$ varies slowly with temperature compared to the exponential decay rates, we obtain the algebraic equation for $\beta/H$:
\begin{align}
    \frac{\beta}{H}
    &=\frac{(1-N_\mathrm{PBH}/N_{\mathrm{tot}})\Gamma_0\,\beta_0/H + (N_\mathrm{PBH}/N_{\mathrm{tot}})\Gamma_\mathrm{PBH}\,\beta_\mathrm{PBH}/H}{(1-N_\mathrm{PBH}/N_{\mathrm{tot}})\Gamma_0 + (N_\mathrm{PBH}/N_{\mathrm{tot}})\Gamma_\mathrm{PBH}}.
    \label{eq:averaged_betaH}
\end{align}
This formula interpolates between the slow homogeneous transition and the fast catalyzed transition depending on the PBH abundance. The behavior of the resulting $\beta/H$ is illustrated in Fig.~\ref{Fig:betaH_fPBH}, where we express the PBH abundance in terms of the fraction $f_\mathrm{PBH}$. The averaged inverse duration $\beta/H$ is bounded from $\beta_0/H$ to $\beta_\PBH/H$. Moreover, the PBH-catalysis effect is highly efficient: even for a very small PBH fraction,
\(f_{\rm PBH}\sim \mathcal{O}(10^{-13})\), the averaged inverse duration parameter \(\beta/H\)
already approaches its upper limit \(\beta_{\rm PBH}/H\), as shown in Fig.~\ref{Fig:betaH_fPBH}. Such result implies that we should carefully consider the catalysis effect, especially in the case of ignoring the PBHs' influence.

In summary, comparing $\beta/H$ with and without PBH demonstrates that PBHs can substantially accelerate the chiral phase transition, particularly for lighter black holes whose near-horizon curvature effects are stronger. Additionally, the PBH-catalysis effect can be considerable even for small PBH fraction. In the following, we discuss how this catalysis influences the evaluation of the resulting GW spectrum.

\subsection{Gravitational-wave spectrum}
\label{subsec:GW spectrum}

First-order phase transitions proceed via the nucleation and subsequent expansion of bubbles of the true vacuum. GWs are generated when these bubbles interact with the surrounding plasma and with each other. 
The dominant GW sources are (i) collisions of bubble walls (or scalar-field shells)~\cite{Witten:1984rs,Kosowsky:1991ua,Kosowsky:1992vn,Kamionkowski:1993fg,Huber:2008hg,Cutting:2018tjt,Cutting:2020nla,Lewicki:2020jiv,Ellis:2020nnr,Lewicki:2020azd,Lewicki:2022pdb,Wang:2025eee}, (ii) long-lived sound waves in the thermal plasma~\cite{Hogan:1986qda, Hindmarsh:2013xza, Hindmarsh:2015qta, Hindmarsh:2017gnf, Jinno:2017fby}, and (iii) MHD turbulence~\cite{Kamionkowski:1993fg, Kosowsky:2001xp, Dolgov:2002ra, RoperPol:2022iel}. Their relative importance is controlled by the microphysics of the transition, in particular by the strength of the transition and by the amount of friction exerted by the plasma on the bubble walls.

In the three-flavor NJL model, the $\sigma$ field strongly couples with the remaining degrees of freedom. The friction exerted by the plasma on the bubble walls is therefore non-negligible, which already points towards a non-runaway scenario~\cite{Caprini:2015zlo,Li:2023xto,Wang:2023lam}. In such a case, the bubble wall velocity approaches a terminal value $v_b < 1$ rather than accelerating all the way to the speed of light. Most of the released energy is then transferred to the bulk fluid motion rather than to scalar field gradients, so that the contribution from scalar-shell collisions is subdominant, while sound waves and turbulence constitute the leading GW source~\cite{Caprini:2015zlo,Helmboldt:2019pan}. In the following, we therefore focus on the sound-wave and turbulence contributions as a good approximation to the total GW spectrum.

\vspace{0.3em}
\noindent\textbf{Phase transition strength parameter \boldmath$\alpha$.}%
\quad
In addition to the nucleation temperature $T_n$ and the inverse duration $\beta/H$, the GW spectrum depends sensitively on the latent heat, denoted by \(\alpha\), which is defined as
\begin{align}
    \alpha(r)
    = \frac{1}{\rho_{\mathrm{rad}}(T_n)}
      \left[
        \Delta V_{\mathrm{eff}}(T_n,r)
        - T_n
          \left.
          \frac{\partial \Delta V_{\mathrm{eff}}(T,r)}{\partial T}
          \right|_{T = T_n}
      \right],
    \label{eq:alpha-def}
\end{align}
where
\begin{align}
    \Delta V_{\mathrm{eff}}(T,r)
    = V_{\mathrm{eff}}(\sigma_f;T,r)
      - V_{\mathrm{eff}}(\sigma_t;T,r)
\end{align}
is the potential difference between the false vacuum $\sigma_f$ and the temperature-dependent global minimum $\sigma_t$. The radiation energy density $\rho_{\mathrm{rad}}(T)$ is given by
\begin{align}
    \rho_{\mathrm{rad}}(T)
    = \frac{\pi^2}{30}\,g_*(T)\,T^4,
\end{align}
with $g_*(T)$ denoting the effective degrees of freedom in the thermal plasma.

The PBHs influence $\alpha(r)$ through the local effective potential $V_{\mathrm{eff}}(\sigma;T,r)$, which reduces the potential barrier and alters the energy released near the event horizon. 
However, for the parameters used in Fig.~\ref{Fig:betaH}, our calculations show that outside the PBH horizon, $\alpha(r)$ remains nearly constant and equals its flat-spacetime value. Therefore, we use the flat-spacetime value of $\alpha$ in calculating the GW spectrum.

\vspace{0.3em}
\noindent\textbf{Sound-wave GW spectrum.}%
\quad
With the above assumptions (non-runaway bubbles, sound-wave dominated signal), the present-day GW spectrum sourced by sound waves can be written as~\cite{Caprini:2015zlo,Hindmarsh:2017gnf}
\begin{align}
    \Omega_{\mathrm{sw}}(f) h^2
    \simeq 2.65 \times 10^{-6}\,
    \left(\frac{H}{\beta}\right)
    \left(\frac{\kappa_v\alpha}{1 + \alpha}\right)^2
    \left(\frac{100}{g_*(T_\star)}\right)^{1/3}
    v_b\, S_{\mathrm{sw}}(f),
    \label{sec4:Omega-sw}
\end{align}
where $h$ is the reduced Hubble parameter, $v_b$ is the terminal bubble-wall velocity, and $\kappa_v$ denotes the efficiency factor describing the fraction of released energy transferred into the bulk fluid motion. 
The dimensionless spectral shape function $S_{\mathrm{sw}}(f)$ and the peak frequency $f_{\mathrm{sw}}$ are given by
\begin{align}
    S_{\mathrm{sw}}(f)
    &= \frac{(f/f_{\mathrm{sw}})^3}
       {\left[1 + \frac{3}{4}\,(f/f_{\mathrm{sw}})^2\right]^{7/2}},
    \label{sec4:S-sw}
    \\
    f_{\mathrm{sw}}
    &\simeq 1.9 \times 10^{-5}\,\mathrm{Hz}\;
       \frac{1}{v_b}\,
       \left(\frac{\beta}{H}\right)
       \left(\frac{T_\star}{100~\mathrm{GeV}}\right)
       \left(\frac{g_*(T_\star)}{100}\right)^{1/6} \,, 
    \label{sec4:f-sw}
\end{align}
where $T_{\star}$ is the reheating temperature defined by $T_{\star} = \left[ 1 + (1 - \kappa_{v})\alpha \right]^{1/4} T_{n}$~\cite{Leitao:2015fmj,Cai:2017tmh}. 
We have numerically confirmed that the approximate $T_n \simeq T_\star$ can be used safely in evaluating the GW spectrum.
For the efficiency factor $\kappa_v$, we use the following approximate fit~\cite{Espinosa:2010hh}
\begin{align}
    \kappa_v
    \simeq \frac{\alpha}{0.73 + 0.083 \sqrt{\alpha} + \alpha} \,.
    \label{sec4:kappa-v}
\end{align}

\vspace{0.3em}
\noindent\textbf{MHD turbulence}%
\quad
In the non-runaway regime, a sub-leading contribution to the stochastic GW background can arise from MHD turbulence in the plasma. A commonly used fit for the present-day GW spectrum from turbulence reads~\cite{Caprini:2015zlo,Ellis:2018mja}
\begin{align}
    \Omega_{\mathrm{tb}} h^2(f)
    \simeq 3.35\times 10^{-4}\,
    \left(\frac{H}{\beta}\right)\,
    \left(\frac{\kappa_{\mathrm{tb}}\,\alpha}{1+\alpha}\right)^{3/2}
    \left(\frac{100}{g_*(T_\star)}\right)^{1/3}
    v_b\, S_{\mathrm{tb}}(f) \,, 
    \label{eq:Omega-tb}
\end{align}
where $\kappa_{\mathrm{tb}}$ denotes the efficiency factor for transferring released vacuum energy into turbulent bulk motion.
The spectral shape $S_{\mathrm{tb}}(f)$ and the peak frequency $f_{\mathrm{tb}}$ are given by~\cite{Caprini:2015zlo,Ellis:2018mja}
\begin{align}
    S_{\mathrm{tb}}(f)
    &= \frac{(f/f_{\mathrm{tb}})^3 \left(1+f/f_{\mathrm{tb}}\right)^{-11/3}}
    {1+8\pi f/h_*},\\
    f_{\mathrm{tb}}
    &\simeq 2.7\times 10^{-5}\,\mathrm{Hz}\;
    \frac{1}{v_b}\left(\frac{\beta}{H}\right)
    \left(\frac{T_\star}{100~\mathrm{GeV}}\right)
    \left(\frac{g_*(T_\star)}{100}\right)^{1/6} \,, 
    \label{eq:Stb-ftb}
\end{align}
where $h_*$ denotes the red-shifted Hubble rate, which is given by~\cite{Kamionkowski:1993fg}
\begin{align}
    h_*
    \simeq 16.5\times 10^{-3}\,\mathrm{Hz}\;
    \left(\frac{T_\star}{100~\mathrm{GeV}}\right)
    \left(\frac{g_*(T_\star)}{100}\right)^{1/6}.
    \label{eq:hstar}
\end{align}
Numerical simulations typically indicate that only a small fraction of the bulk kinetic energy becomes turbulent, often parametrized as $\kappa_{\mathrm{tb}}=\epsilon\,\kappa_v$ with $\epsilon\simeq 0.05$~\cite{Caprini:2015zlo,Ellis:2018mja}. Under this assumption, the turbulence contribution is usually subdominant compared to sound waves, and the total stochastic background is well approximated by the sound-wave signal alone.

\begin{figure}[t]
    \centering
    \begin{subfigure}{0.48\textwidth}
        \centering
        \includegraphics[width=\linewidth]{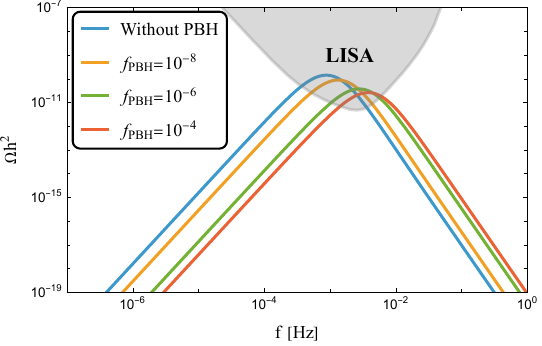}
        \caption{$\Lambda = 1~\mathrm{TeV}$}
        \label{Fig:Omegah2L1000}
    \end{subfigure}
    \hfill
    \begin{subfigure}{0.48\textwidth}
        \centering
        \includegraphics[width=\linewidth]{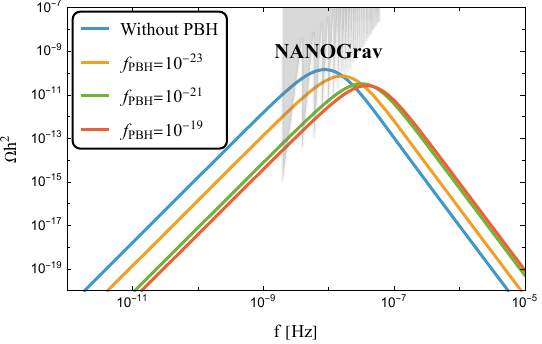}
        \caption{$\Lambda = 10~\mathrm{MeV}$}
        \label{Fig:Omegah2L0.01}
    \end{subfigure}
    \caption{
        Gravitational-wave spectra predicted for the benchmark
        $G_S \Lambda^2 = 8$,
        $G_A \Lambda^5 = -448$ and
        $m_0 \Lambda^{-1} = 10^{-3}$,
        assuming a bubble-wall velocity $v_b = 0.9$.
        In both panels, the blue curve corresponds to the flat-spacetime case, while the orange, green and red curves show the GW spectra for each PBH fraction.
        In panel~(a), $\Lambda = 1~\mathrm{TeV}$ yields a peak in the milli-Hz band, with the gray shaded region indicating the target sensitivity range of LISA.
        In panel~(b), we take $\Lambda = 10~\mathrm{MeV}$. 
        In this case, the GW peak is within the nano-Hz band, overlapping with the sensitivity region of pulsar-timing arrays such as NANOGrav, illustrated by the gray band.
        In both cases, the PBH background slightly suppresses the peak amplitude and shifts the peak to higher frequencies, but the signal remains within the reach of the corresponding GW experiments.
    }
    \label{Fig:Omegah2}
\end{figure}

For illustration, we take $v_b = 0.9$ as a representative relativistic value and evaluate eqs.~\eqref{sec4:Omega-sw}–\eqref{sec4:kappa-v} for the benchmark point given in eq.~\eqref{sec3:BenchMark}. 
The resulting GW spectra are shown in Fig.~\ref{Fig:Omegah2} for the flat spacetime and several values of the PBH fraction $f_{\rm PBH}$. 
As discussed in Sec.~\ref{subsec:PBH-catalysis}, the presence of PBHs increases $\beta/H$ compared to that in flat-spacetime, while leaving $\alpha$ hardly changed.
According to eq.~\eqref{sec4:Omega-sw}, this leads to a moderate suppression of the peak amplitude and, via Eq.~\eqref{sec4:f-sw}, to a shift of the peak frequency towards higher values. 
Both features are clearly visible in Fig.~\ref{Fig:Omegah2}. 
The large $f_{\rm PBH}$ makes the peak frequency higher than the flat spacetime result. 
In addition, the peak amplitude decreases slightly as $f_{\rm PBH}$ is large. 
These results imply that the existence of PBHs may be seriously taken into account in discussing implications of GWs on physics beyond the standard model.  

As shown in Fig.~\ref{Fig:Omegah2}, depending on the cutoff scale $\Lambda$, the SGWB of the chiral phase transition can potentially be probed by space-based interferometers in the milli-Hz band ($\Lambda = 1\,{\rm TeV}$) or by pulsar timing arrays in the nano-Hz band ($\Lambda = 10\,{\rm MeV}$). The PBH background mainly reshapes the spectrum by shifting the peak frequency and reducing the amplitude by a factor of a few, but does not push the signal out of the observable windows of LISA or NANOGrav for the benchmarks considered here. 
This finding implies that the small amount of PBHs can drastically affect the phase transition dynamics and the resulting GW spectra. 
We note that the scenario with $\Lambda = 10\,{\rm MeV}$ is expected to be strongly constrained by the cosmic microwave background due to the reheating effect and the residual particles in the dark sector~\cite{Bai:2021ibt}. 
Since our main purpose is to illustrate how the presence of a small amount of PBH can significantly influence dark chiral phase transitions at such low energy scales, we do not discuss the details on the cosmological constraints.

\section{Discussions \label{sec:discussions}}

Here we give several comments on our analysis and the implications of our finding. 

\paragraph{Implications for Hawking emission channels}

The emergence (or absence) of a symmetry-restored layer outside the event horizon may affect the effective set of light degrees of freedom relevant for Hawking emission, and thereby the PBH mass loss rate $P_{\rm PBH}$.
In the standard description, the total power radiated by a non-rotating black hole is~\cite{Page:1976df}
\begin{align}
    P_{\rm PBH} \equiv -\frac{\md M_{\rm PBH}}{\md t}
    = \sum_i \int_0^\infty \frac{\md \omega}{2\pi}\,
    \frac{\omega\,\Gamma_i(\omega)}{\exp(\omega/T_{\rm BH}) \pm 1},
\end{align}
where $\Gamma_i(\omega)$ is the graybody factor for species $i$ and $T_{\rm BH}=1/(4\pi r_s)$.
The key point is that the set of species that contribute efficiently depends on which degrees of freedom are effectively light near the event horizon~\cite{Flachi:2015fna}.

\begin{figure}[t]
    \centering
    \begin{subfigure}{0.48\textwidth}
        \centering
        \includegraphics[width=\linewidth]{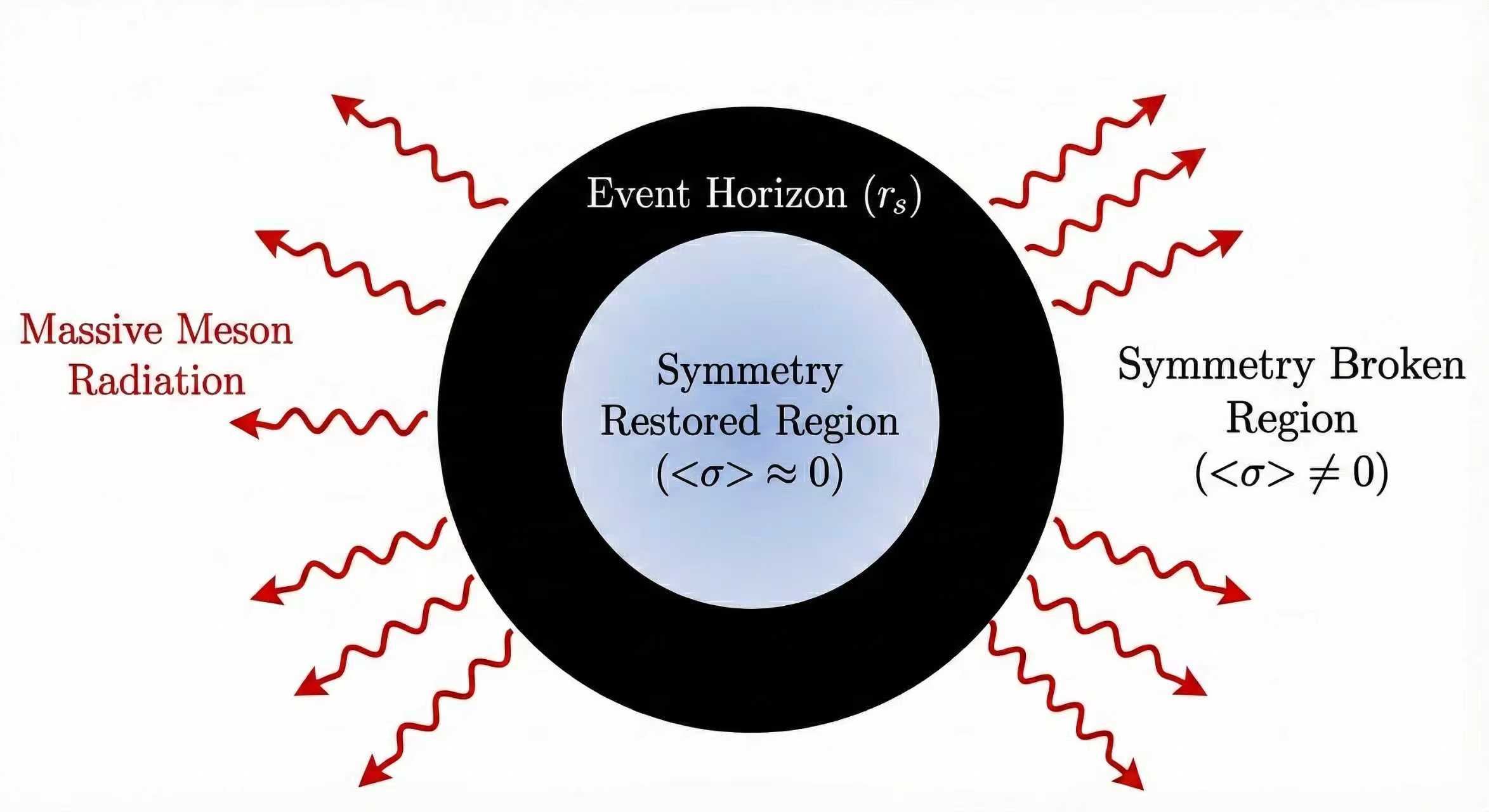}
        \caption{Hadronic degrees of freedom dominate}
        \label{Fig:Hawking Radiation meson}
    \end{subfigure}
    \hfill
    \begin{subfigure}{0.48\textwidth}
        \centering
        \includegraphics[width=\linewidth]{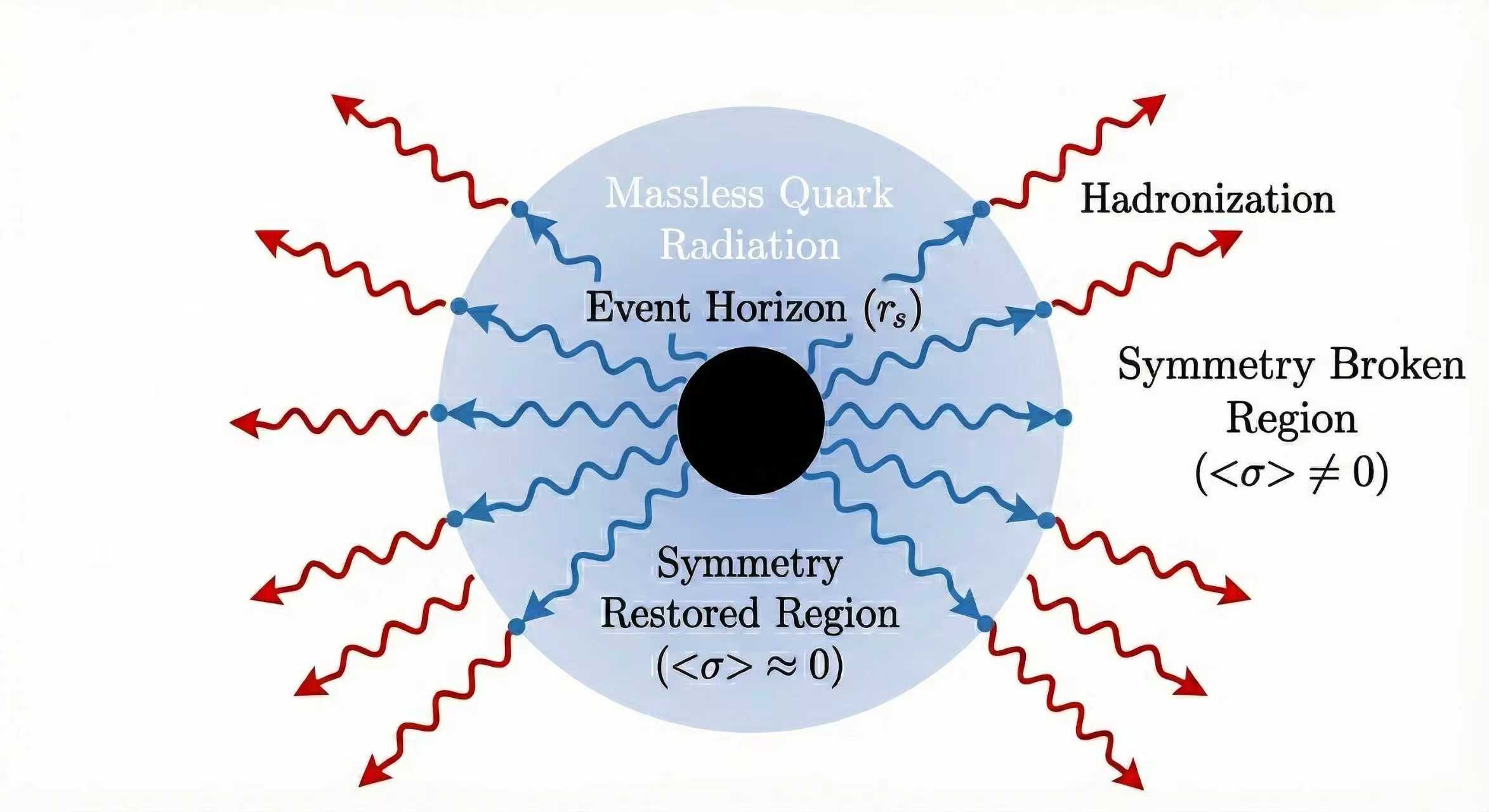}
        \caption{Deconfined quarks and gluons dominate}
        \label{Fig:Hawking Radiation quark}
    \end{subfigure}
    \caption{
        Schematic Hawking emission near the PBH horizon in the absence (left) or presence (right)
        of a symmetry-restored shell outside the horizon.
    }
    \label{Fig:Hawking radiation}
\end{figure}

If the symmetry-restored (deconfined) region is entirely hidden behind the event horizon, as shown in Fig.~\ref{Fig:Hawking Radiation meson}, the exterior remains in the symmetry-broken phase and the relevant low-energy excitations are hadronic.
For sufficiently low $T_{\rm BH}$, heavy states are Boltzmann suppressed, which corresponds to a smaller effective coefficient
$\alpha_{\rm eff}$ in the commonly used parametrization $-\md M_{\rm PBH}/\md t \simeq \alpha_{\rm eff}/M_{\rm PBH}^2$~\cite{MacGibbon:1990zk,MacGibbon:1991tj}.

Conversely, if a symmetry-restored shell extends outside the even horizon, as in Fig.~\ref{Fig:Hawking Radiation quark},
quarks and gluons act as the relevant light degrees of freedom around the even horizon. 
Thus, the effective number of radiated species is increased and the evaporation power may be enhanced.
The resulting hadronic final states can be modeled via QCD jet fragmentation once the partons propagate into the symmetry-broken region~\cite{MacGibbon:1990zk,MacGibbon:1991tj}.
Although these partons will hadronize, such hadronization mainly redistributes the emitted energy into hadronic jets
and does not change the horizon-level power controlling $-\md M_{\rm PBH}/\md t$ unless strong reprocessing effects develop.
This parallels the standard treatment of QCD jet emission in PBH evaporation and related discussions of QED/QCD photosphere formation~\cite{Heckler:1996vg,MacGibbon:2008uf}.

\paragraph{The effect from Hawking radiation}

We have analyzed the combined effects of thermal corrections and PBH curvature on the chiral symmetry breaking. 
In general, when the PBHs exist in thermal plasma, the temperature around the PBHs is modified due to the Hawking radiation and accretion effects. 
However, as confirmed in Ref.~\cite{He:2024wvt}, the plasma near the BH cannot be in equilibrium with Hawking radiations because of the compactness of BH.
Thus, we expect that the thermal correction to the effective potential mainly comes from the thermal plasma contribution. 
Hawking emission can, in principle, heat the plasma in the vicinity of the PBH~\cite{He:2024wvt}. If local thermal equilibrium were established, the plasma temperature near the event horizon would approach the Hawking temperature $T_{\mathrm{BH}}$, as assumed in some previous studies of chiral phase transitions around PBHs~\cite{Flachi:2011sx}. However, the extreme compactness of PBHs implies that the mean free path of Hawking radiation is much larger than the PBH radius, so local thermal equilibrium is not achieved in the region $r \sim r_s$~\cite{He:2024wvt}. 
Actually, when the BH heating is treated consistently, the resulting net temperature increase of the plasma, \(\Delta T_{\mathrm{plasma}}\), is found to be negligible compared to the cosmological plasma temperature \(T_{\rm plasma}\) for the PBH mass considered in our work, corresponding to \(r_s\Lambda\sim \mathcal{O}(1)\)~\cite{He:2024wvt},
\begin{align}
    \frac{\Delta T_{\mathrm{plasma}}}{\Lambda}
    \sim \mathcal{O}(10^{-6}) \ll \frac{T_{\rm plasma}}{\Lambda}.
\end{align}
We therefore regard $T$ in the effective potential as the cosmological plasma temperature and treat the PBH background as a purely gravitational perturbation of the chiral dynamics.

\paragraph{Baby primordial black hole formation via parent PBH catalysis effects}

We give a comment on the baby PBH formation via the parent PBH formation. 
As shown in Fig.~\ref{Fig:betaH_fPBH}, the existence of the PBHs can enhance the bubble nucleation rate. 
In other words, the phase transition is delayed in the domain without the PBHs. 
Thus, the inhomogeneous PBH distribution may cause the baby PBH formation as pointed out in Ref.~\cite{Jinno:2023vnr}. 
This intriguing possibility is the subject of future work.

\section{Conclusions}
\label{sec:conclusions}

In this work, we have studied the impact of PBHs on the chiral phase transition and the associated SGWB using the three-flavor NJL model as an effective field theory. 
Our analysis includes the construction of the NJL effective potential in curved spacetime, the estimation of the bubble nucleation rate in a Schwarzschild background, and the analysis of the prediction on the expected GW spectra. 
The main results are summarized as follows:

\begin{itemize}

    \item We derived the finite-temperature NJL effective potential in a generic curved spacetime using the Riemann normal coordinate expansion. Specializing to the Schwarzschild geometry, we obtained a compact decomposition of the PBH-modified potential shown in eq.~\eqref{sec3:effective potential-1}.
    This framework clarifies that the curvature effects have a significant impact on chiral symmetry breaking.

    \item We discovered a novel local phase structure in the vicinity of a PBH based on this effective potential. 
    While the NJL model in the flat spacetime exhibits a conventional first-order chiral phase transition, the system can undergo a re-entrant sequence near the PBH horizon: a second-order transition to a broken phase, a first-order transition between two distinct broken phases, and the symmetry restoration.
    This three-stage pattern, driven purely by curvature effects, represents a new type of chiral phase structure around the PBHs.

    \item We quantified the PBH-induced catalysis by obtaining the \(O(3)\) bounce solution in the Schwarzschild spacetime background and derived the inverse duration parameter \(\beta/H\) depending on the PBH fraction in eq.~\eqref{eq:averaged_betaH}. 
    Compared to the flat spacetime case, we found that a large PBH fraction significantly enhances \(\beta/H\), while the strength parameter \(\alpha\) remains close to its flat spacetime value. 
    In this sense, PBHs act as efficient catalysts without significantly altering the latent heat of the phase transition.

    \item We translated the PBH-induced modifications of $\beta/H$ and $\alpha$ into predictions for the GW spectrum, focusing on the sound-wave contribution in the non-runaway regime.
    For the TeV-scale benchmark, the resulting GW spectrum has a peak in the milli-Hz band, where PBHs shift the peak frequency from sub-milli-Hz to a few milli-Hz and suppress the peak amplitude by a factor of a few, in accordance with the standard scaling with $\beta/H$. 
    For a low-scale benchmark with $\Lambda \sim \mathcal{O}(\mathrm{MeV})$, the same mechanism shifts the signal into the nano-Hz band, placing it within the sensitivity range of pulsar-timing arrays such as NANOGrav. 
    In both cases, there can be $\mathcal{O}(1)$ deviations in the peak frequency and amplitude of SGWBs between the case with and without the PBHs. 
    
\end{itemize}

Our study provides a general template for analyzing phase transitions in curved spacetime. 
In particular, the calculation framework for the typical phase transition duration given in eq.~\eqref{eq:averaged_betaH} can be applied to other phase transition models, alternative PBH mass scales, and realistic PBH mass functions. 
We leave these interesting studies as future works.

\section*{Acknowledgment}

M.T. is grateful to Takumi Kuwahara for fruitful discussions. J.-C.W. and J.-J.Z would like to appreciate Banghui Hua and Jiang Zhu for enlightening discussions.

\appendix

\section{Derivation of the effective potential}\label{appA}

Substituting eq.~\eqref{sec3:Coefficients} and eq.~\eqref{sec2:G in k space} into eq.~\eqref{eq:Veff_flat}, the zero-temperature effective potential in curved spacetime can be written as
\begin{align}\label{appA:Veff-zeroT}
    V_\mathrm{eff}(\sigma;T=0,r)
    &= \frac{3}{8 G_S}\sigma^{2}
    - \frac{G_A}{16 G_S^{3}}\sigma^{3}
    - i \int_{M(0)}^{M(\sigma)} \md s\, s
      \int \frac{\md^{4}k}{(2\pi)^{4}}
      \left[
        \frac{4}{k^{2}-s^{2}}
        - \frac{7 r_s^{2}}{5 r^{6}} \frac{s^{2}}{(k^{2}-s^{2})^{4}}
      \right].
\end{align}

Using the finite-temperature replacement rules in eq.~\eqref{sec2:Feynman rules in FTFT}, eq.~\eqref{appA:Veff-zeroT} becomes
\begin{align}\label{appA:Veff-finiteT}
    V_\mathrm{eff}(\sigma;T,r)
    &= \frac{3}{8 G_S}\sigma^{2}
    - \frac{G_A}{16 G_S^{3}}\sigma^{3}
    -  \int_{M(0)}^{M(\sigma)} \md s\, 4 s\, T
      \sum_{n=-\infty}^{\infty}
      \int \frac{\md^{3}k}{(2\pi)^{3}}
      \frac{1}{\omega_n^{2}+\omega_k^{2}} \notag\\
    &\quad
    -  \int_{M(0)}^{M(\sigma)} \md s\,
      \frac{7 s^{3}\, T\, r_s^{2}}{5 r^{6}}
      \sum_{n=-\infty}^{\infty}
      \int \frac{\md^{3}k}{(2\pi)^{3}}
      \frac{1}{(\omega_n^{2}+\omega_k^{2})^{4}},
\end{align}
where $\omega_n \equiv (2n+1)\pi T$ is the fermionic Matsubara frequency and $\omega_k \equiv \sqrt{\boldsymbol{k}^{\,2}+s^{2}}$.

To relate the curvature term to derivatives acting on the basic thermal integral, we use the identity
\begin{align}\label{appA:derivative-trick}
    -\frac{1}{6}\left(\frac{\partial}{\partial s^{2}}\right)^{3}
    \frac{1}{\omega_n^{2}+\omega_k^{2}}
    = \frac{1}{(\omega_n^{2}+\omega_k^{2})^{4}}.
\end{align}
Furthermore, the standard Matsubara sum yields
\begin{align}\label{appA:matsubara-sum}
    T \sum_{n=-\infty}^{\infty}\frac{1}{\omega_n^{2}+\omega_k^{2}}
    = \frac{1}{\omega_k}
      \left(
        \frac{1}{2}
        - \frac{1}{e^{\omega_k/T}+1}
      \right).
\end{align}
Inserting eq.~\eqref{appA:derivative-trick} and eq.~\eqref{appA:matsubara-sum} into eq.~\eqref{appA:Veff-finiteT}, the effective potential can be rearranged into the four contributions shown in eq.~\eqref{sec3:effective potential-1}. As clarified in the main text, each contribution can be identified with either the zero- or finite-temperature part and with either the flat-spacetime or the PBH-induced curvature contribution.

\section{Meson-field propagator}\label{app:meson-propagator}

The one-loop inverse propagator of the $\sigma$ meson in flat spacetime is given by~\cite{Helmboldt:2019pan}
\begin{align}
    \Gamma_{\sigma\sigma}(p^2,\sigma)
    &= -\frac{3}{4G}
    + \frac{3G_D\,\sigma}{8G^3}
    - \left(1-\frac{G_D \sigma}{4G^2}\right)^{\!2} 3N_c\, I_S(p^2,\sigma)
    + \frac{G_D}{G^2}\,3N_c\, I_V(\sigma),
    \label{app:inverse propagator}
\end{align}
where the loop integrals are
\begin{align}
    \begin{aligned}
        I_V(\sigma)
        &= \int\frac{\md^4k}{i(2\pi)^4}\,
        \frac{M(\sigma)}{k^2-M(\sigma)^2}, \\
        I_S(p^2,\sigma)
        &= \int\frac{\md^4k}{i(2\pi)^4}\,
        \frac{\Tr\!\left[(\slashed{k}+\slashed{p}+M(\sigma))(\slashed{k}+M(\sigma))\right]}
        {\big((k+p)^2-M(\sigma)^2\big)\big(k^2-M(\sigma)^2\big)}.
    \end{aligned}
\end{align}
At finite temperature, one replaces the loop integral according to the replacement rule in eq.~\eqref{sec2:Feynman rules in FTFT}.

The (finite-temperature) wave-function renormalization is defined as
\begin{align}
    Z_{\sigma}^{-1}(\sigma)
    = \left.\frac{\md \Gamma_{\sigma\sigma}\!\left(p^{2},\sigma\right)}{\md p^{2}}\right|_{p^{2}=0}.
    \label{app:Zsigma-def}
\end{align}
Expanding the inverse propagator around $p^2=0$,
\begin{align}
    \Gamma_{\sigma\sigma}(p^2,\sigma)
    = \Gamma_{\sigma\sigma}(0,\sigma) + Z_\sigma^{-1}(\sigma)\,p^2 + \cdots,
\end{align}
one can recover the canonical kinetic term by the field redefinition $\sigma_R=Z_\sigma^{-1/2}\sigma$.
In the main text, we continue to work with $\sigma$ and the non-canonical kinetic prefactor $Z_\sigma^{-1}(\sigma)$.

Substituting eq.~\eqref{app:inverse propagator} into eq.~\eqref{app:Zsigma-def}, one obtains~\cite{Helmboldt:2019pan}
\begin{align}
    Z_\sigma^{-1}(\sigma)
    = -3N_c\left(1-\frac{G_D}{4G^2}\sigma\right)^{\!2}
    \left[
        -2A_0+2B_0+8C_0
        -2\ell_A(u)+2\ell_B(u)+8\ell_C(u)
    \right],
\end{align}
with
\begin{align}
    \begin{aligned}
        A_0
        &= \frac{1}{16\pi^2}\left[
            \log\left(1+\frac{\Lambda^2}{M(\sigma)^2}\right)
            -\frac{\Lambda^2}{\Lambda^2+M(\sigma)^2}
        \right],\\
        B_0
        &= -\frac{1}{32\pi^2}\frac{\Lambda^4}{\big(M(\sigma)^2+\Lambda^2\big)^2},
        \qquad
        C_0
        = \frac{1}{96\pi^2}\frac{3M(\sigma)^2\Lambda^4+\Lambda^6}{\big(M(\sigma)^2+\Lambda^2\big)^3},
    \end{aligned}
\end{align}
and dimensionless thermal variable $u \equiv \frac{|M(\sigma)|}{T}$. 
The thermal functions read
\begin{align}
    \ell_A(u)
    &= -\frac{1}{4\pi^2}\int_0^\infty \md x \Bigg(
        \frac{x^2}{(\sqrt{x^2 + u^2})^3} \frac{1}{1 + \exp (\sqrt{x^2 + u^2})}
        + \frac{1}{2} \frac{x^2}{(\sqrt{x^2 + u^2})^2} \frac{1}{1 + \cosh (\sqrt{x^2 + u^2})}
    \Bigg), \\
    \ell_B(u)
    &= \frac{u^2}{16\pi^2}\int_0^\infty \md x \Bigg(
        \frac{3x^2}{(\sqrt{x^2 + u^2})^5} \frac{1}{1 + \exp (\sqrt{x^2 + u^2})}
        + \frac{3x^2}{2(\sqrt{x^2 + u^2})^4} \frac{1}{1 + \cosh (\sqrt{x^2 + u^2})}\notag\\
        &\quad+ \frac{x^2}{2(\sqrt{x^2 + u^2})^5} \frac{1}{1 + \cosh (\sqrt{x^2 + u^2})}
    \Bigg), \\
    \ell_C(u)
    &= -\frac{u^4}{96\pi^2}\int_0^\infty \md x \Bigg(
        \frac{15x^2}{(\sqrt{x^2 + u^2})^7} \frac{1}{1 + \exp (\sqrt{x^2 + u^2})}
        + \frac{15x^2}{2(\sqrt{x^2 + u^2})^6} \frac{1}{1 + \cosh (\sqrt{x^2 + u^2})} \notag \\
        &\quad + \frac{3x^2}{(\sqrt{x^2 + u^2})^5} \frac{\tanh(\sqrt{u^2 + x^2}/2)}{1 + \cosh( \sqrt{x^2 + u^2})}
        + \frac{x^2}{2(\sqrt{x^2 + u^2})^4} \frac{1}{1 + \cosh (\sqrt{x^2 + u^2})} \notag \\
        &\quad - \frac{3x^2}{2(\sqrt{x^2 + u^2})^4} \frac{1}{\left(1 + \cosh (\sqrt{x^2 + u^2})\right)^2}
    \Bigg).
\end{align}

\newpage

\bibliography{ref.bib}{}
\bibliographystyle{utphys28mod}

\end{document}